\documentclass[10pt,twocolumn]{article} 

\usepackage{times}
\usepackage{graphicx}
\usepackage{amssymb}
\usepackage{url,hyperref}
\usepackage{amsmath}

\begin{document}

\title{\bf Interferometric Autocorrelation Measurements of Supercontinuum based on Two-Photon Absorption}

\author{Shanti Toenger$^1$, Roosa M{\"a}kitalo$^1$, Jani Ahvenj{\"a}rvi$^1$, Piotr Ryczkowski$^1$, Mikko N{\"a}rhi$^1$,\\ John M. Dudley$^2$ and Go{\"e}ry Genty$^{1,*}$ \vspace{0.3cm} \\ 
$^1$\small Laboratory of Photonics, Tampere University of Technology, Tampere, Finland\\
$^2$\small Institut FEMTO-ST, UMR 6174 CNRS-Universit{\'e} Bourgogne Franche-Comt{\'e}, Besan\c{c}on, France\\
*\small Corresponding author: goery.genty@tut.fi}

\maketitle
\thispagestyle{empty}

\begin{abstract}
We report on interferometric autocorrelation measurements of broadband supercontinuum light in the anomalous dispersion regime using two-photon absorption in a GaP photodetector. The method is simple, low-cost, and provides a direct measure of the second-order coherence properties, including quantitative information on the coherence time and average duration of the supercontinuum pulses as well as on the presence of temporally coherent sub-structures. We report measurements in regimes where the supercontinuum is coherent and incoherent. In the former case, the interferometric measurements are similar to what is observed for mode-locked laser pulses while in the latter case the interferometric measurements and coherence properties are shown to have characteristics similar to that of a stationary chaotic light source.
\end{abstract}

\section{Introduction}

Supercontinuum (SC) generation in nonlinear fibers has attracted significant attention because of its potential for a wide range of applications. Depending on the intended application, the coherence properties of SC need to be carefully considered and characterized. For example, applications in metrology, telecommunications, or coherent spectroscopy require high shot-to-shot phase stability and long coherence time, while applications in imaging and sensing may be more tolerant to fluctuations and do not require SC with especially long coherence time. Several studies of SC coherence properties have been conducted, both theoretically and experimentally \cite{Dudley2002,Gu2003b,Gu2002}. In these studies, the quantification of the coherence is typically realized utilizing the second-order (in fields) normalized degree of spectral coherence \cite{Dudley2002}. This description, however, only characterizes the shot-to-shot phase stability and a more complete characterization of SC coherence may be obtained using correlation functions of the second-order coherence theory of non-stationary light \cite{Genty2010a,Genty2011,Narhi2016a}. 

The second-order degree of temporal coherence is characteristics of the emission mechanism of a light source \cite{MANDEL1965,Perina1985}. For example, the normalized second-order degree of coherence of perfectly classical fields at zero-delay is equal to one for a quasi-monochromatic continuous laser while for a chaotic stationary source it is increased to two \cite{MANDEL1965,Perina1985}. The simplest method for quantifying the second-order degree of coherence of a light source is to use a nonlinear autocorrelation technique such as e.g. intensity autocorrelation (AC) \cite{Diels1978} or interferometric autocorrelation (IAC) \cite{Salehi1990}.  A fundamental difference between AC and IAC is that IAC also provides information about the phase, which is particularly useful when characterizing short pulses \cite{Diels1985,Fork1987}. In addition, AC and IAC measures the second and fourth power of the electric field amplitude, respectively, such that IAC yields a higher contrast and enhanced sensitivity to the pulse shape. Specifically, the envelope of the IAC profile for non-stationary (pulsed) light has a peak-to-background ratio of 8:1, compared to the AC profile with ratio of 2:0. 

Both interferometric or intensity autocorrelation may be performed using second-harmonic generation in a nonlinear crystal \cite{Ippen1984,Diels1985} or via two-photon absorption (TPA) in a semiconductor \cite{Ranka1997,Boitier2009,Cong2018}. In second-harmonic autocorrelation, the beam to be characterized interferes with a delayed replica of itself in a nonlinear crystal that generates light at the second-harmonic frequency of the input light. The autocorrelation trace is then proportional to the second-harmonic intensity measured as a function of delay between the interfering beams. TPA-based IAC on the other hand is more compact, inexpensive and does not require phase-matching as compared to the second-harmonic approach \cite{Takagi1992,Barry1996,Ranka1997}, and indeed characterization of short laser pulses \cite{Hundertmark2004,Chong2014,Shin2016} or photon bunching in a blackbody source \cite{Boitier2009,Boitier2013} with resolution down to the femtosecond level have been reported using this simple approach.  

Second-harmonic based IAC measurements of ultra-short, coherent broadband SC have been reported earlier, including the study of compressed SC in a hollow-core fiber filled with a noble gas \cite{Cavalieri2007} or the study of mid-infrared filamentation and post-compression in bulk dielectrics \cite{Liang2014}. TPA-based IAC measurements of fiber-based SC have also been performed, however they have been restricted to either highly stable and coherent SC generated in the normal dispersion regime \cite{Hooper2011}, or to study relatively narrowband and filtered spectral components \cite{Farrell2012}.  
 
Here, we expand previous studies and report TPA-IAC measurements of broadband supercontinuum using a semiconductor GaP photodetector. In contrast to earlier studies \cite{Hooper2011,Liang2014}, measurements are performed in the anomalous dispersion regime and for the full duration of the SC pulses extending over several picoseconds. We study both the case of a (phase-stable) coherent and incoherent SC. Although IAC does not provide a direct measure of the (average) phase across SC pulses as frequency-resolved optical gating \cite{Trebino1997} or spectral phase interferometry for direct electric field reconstruction \cite{Iaconis1998}, it nevertheless reveals useful information about the presence of coherent sub-structure and provides a direct measure of the second-order coherence properties. Our experimental results are in good agreement with numerical simulations and they show that simple, TPA-based, interferometric autocorrelation measurements can be a convenient means to characterize the temporal properties of broadband supercontinuum light with direct access to the coherence time.

\section{Numerical Simulations}

\begin{figure}[htbp]
\centering
\includegraphics[width=\linewidth]{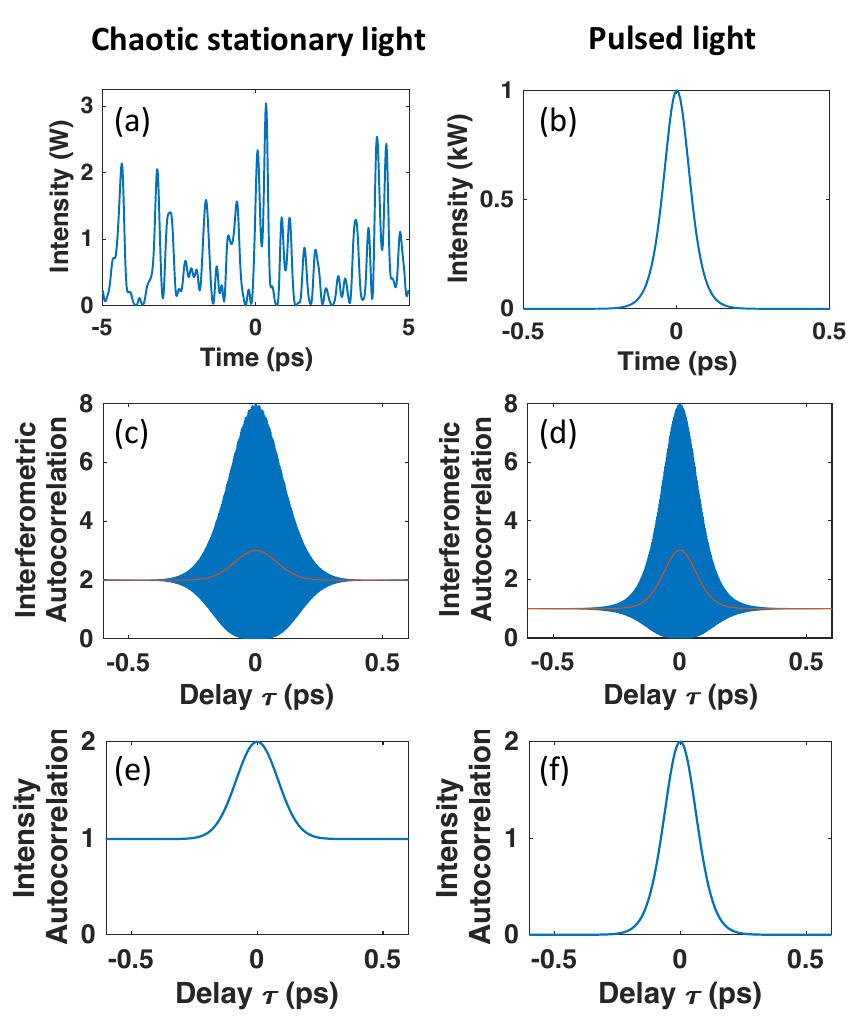}
\caption{Comparison of the properties of stationary chaotic light (left panels) and mode-locked pulses (right panels). (a) and (b): time-domain intensities, (c) and (d): IAC function (blue) and superimposed low-pass filtered trace (red), (e) and (f) AC function. The coherence time of both chaotic light and mode-locked pulses here is $\tau_c 100$ fs.}
\label{fig:Stationary-Pulse}
\end{figure}

The interferometric autocorrelation profile as a function of delay can be written as \cite{Diels2006}
\begin{equation}
I(\tau)=\int_{-\infty}^{\infty}\left|\left[E(t)+E(t-\tau)\right]^{2}\right|^{2}dt.
\end{equation}
Expanding the term in the bracket, the IAC profile consists of four distinct contributions \cite{Diels2006}:
\begin{equation}
I(\tau)=I_{bg}+I_{AC}(\tau)+I_{\omega}(\tau)+I_{2\omega}(\tau)
\end{equation}
where
\begin{align*}
&I_{bg}=2\int_{-\infty}^{\infty}I^{2}(t)\thinspace dt \\
&I_{AC}(\tau)=4\int_{-\infty}^{\infty}I(t)I(t-\tau)\thinspace dt \\
&I_{\omega}(\tau)=4\int_{-\infty}^{\infty}\left[I(t)+I(t-\tau)\right]\Re\left[E(t)E^{*}(t-\tau)\right]\thinspace dt \\
&I_{2\omega}(\tau)=2\int_{-\infty}^{\infty}\Re\left[E^{2}(t)E^{*}{}^{2}(t-\tau)\right]\thinspace dt \\
\end{align*}
correspond to the second harmonic of individual pulses (i.e the background intensity), intensity autocorrelation, field autocorrelation (i.e. the spectrum), and the interferogram of the second harmonic, respectively. Normalizing this equation relative to the background intensity $I_{bg}$, the IAC profile can then be expressed as
\begin{equation} \label{eq:I_IAC}
I_{IAC}(\tau)=1+\frac{I_{AC}(\tau)}{I_{bg}}+\frac{I_{\omega}(\tau)}{I_{bg}}+\frac{I_{2\omega}(\tau)}{I_{bg}}.
\end{equation}
Equation \ref{eq:I_IAC} also shows that the AC function $I_{AC}$ can be obtained from the IAC trace by filtering out the fast oscillating components $I_{\omega}(\tau)$ and $I_{2\omega}(\tau)$ and subtracting the normalized background (i.e 1). The AC and IAC functions for stationary chaotic light and mode-locked pulses are illustrated in Fig. \ref{fig:Stationary-Pulse}. It can be seen that for chaotic stationary light, the peak-to-background ratio of the IAC function is equal to 8:2, while for mode-locked pulses light the ratio is increased to 8:1. The corresponding AC function obtained then yields a ratio of 2:1 for chaotic stationary light and 2:0 for pulsed light. 

\begin{figure}[htbp]
\centering
\includegraphics[width=\linewidth]{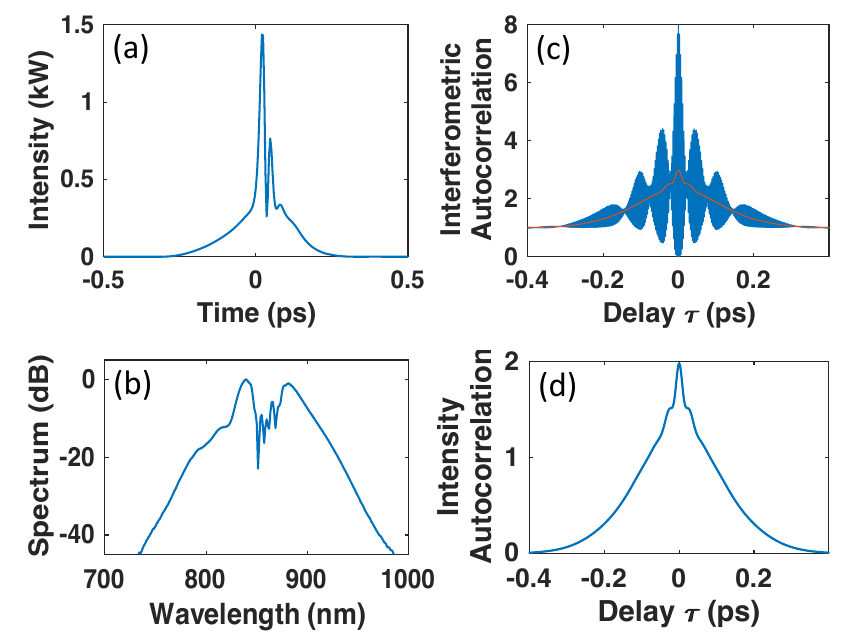}
\caption{Simulated coherent SC with $P_{p} = 420$~W. (a) Intensity in time domain. (b) Normalized spectrum. (c) Interferometric autocorrelation (blue) and superimposed AC function trace obtained by low-pass filtering the IAC profile (red). (d) AC function.}
\label{fig:Sims-Coherent-SC}
\end{figure}

In order to understand the specific features associated with the IAC profile of a broadband supercontinuum, we first perform numerical simulations based on the generalized nonlinear Schrödinger equation \cite{Dudley2006a}. Note that the simulations presented below use the parameters of the experiments reported in the next section. The model includes dispersion coefficients up to the tenth order, the frequency-dependence of the nonlinear response as well as the (experimental) Raman gain of silica \cite{Dudley2006a}. Noise on the input pulse is also included in the form of one-photon with random phase per spectral bin. We first analyze the case where the generating dynamics are fully deterministic with no pulse-to-pulse fluctuation, such that the SC is fully phase-stable and coherent (in any field order). The injected peak power is $P_p = 420$~W corresponding to an input soliton number $N=7$. The temporal intensity and average SC spectrum are shown in Fig. \ref{fig:Sims-Coherent-SC}(a,b). The spectrum extends over a 200~nm span (-30 dB bandwidth) while in the time-domain the dynamics are essentially dominated by higher-order soliton compression and fission with a resulting temporal SC duration of c.a. 100~fs. The corresponding computed IAC and AC traces are shown in Fig. \ref{fig:Sims-Coherent-SC}(c,d). One can see that, the IAC profile (blue line) consists of a slowly-varying envelope including a number of side-lobes with very fast oscillations across the full duration of the envelope. The oscillation period is determined by the central wavelength of the SC pulses, and interference across the full duration of the IAC envelope is an indication of the near perfect coherence property of the SC field. It also provides a direct measure of the coherence time of the SC pulses $\tau_c\approx300$~fs, taken as the delay over which interference are observed underneath the IAC envelope. The total number of side-lobes on one side of the zero-delay corresponds to the number of coherent sub-structures in the temporal intensity profile of the SC pulses. The AC profile that can be extracted from the IAC by simple low-pass filtering on the other hand mostly provides information on the duration of the SC pulses with little details about the presence of these sub-structures. The envelope of the central spike in the AC trace has a contrast of 2:0 with quite similar characteristics to that of mode-locked pulses (see Fig. 1).  

\begin{figure}[htbp]
\centering
\includegraphics[width=\linewidth]{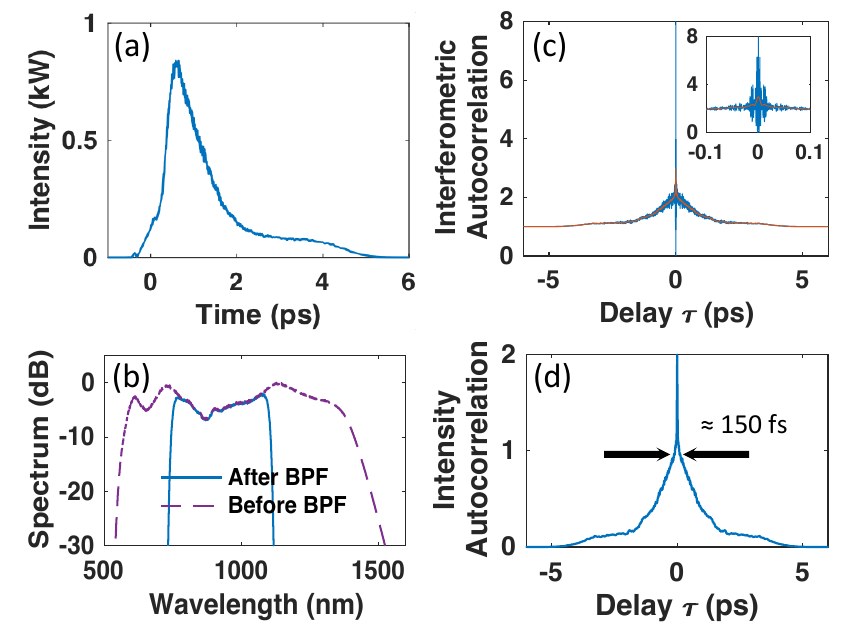}
\caption{Simulated incoherent SC with $P_{p}=18$~kW. (a) and (b) plots the SC average temporal intensity and spectrum computed over 2000 realizations, respectively. (c) Interferometric autocorrelation (blue) with and superimposed AC function trace obtained by low-pass filtering the IAC profile (red) computed over 50 realizations. (d) AC function. Note that for the purpose of comparison with the experiment, a super-Gaussian bandpass filter over the range of 750 nm to 1100 nm was applied. These wavelength range corresponds to the absorption range of the GaP photodetector and long-pass filter used to reduce the linear absorption in the lower wavelength range.}
\label{fig:Sims-Incoherent-SC}
\end{figure}

We next examine the case where the injected peak power is increased to $P_p = 18$~kW corresponding to an input soliton number $N=45$. The spectrum now extends over more than one octave (-30 dB bandwidth) and we note the absence of fine structure in the average (over 2000 distinct realizations) temporal envelope and spectrum of the SC shown in Fig. \ref{fig:Sims-Incoherent-SC}(a,b). The absence of fine structure is a clear indication of large shot-to-shot fluctuations associated with the relatively large soliton number and triggered by noise-seeded modulation instability \cite{Dudley2006a}. For the purpose of comparison with the experiments reported below, a super-Gaussian bandpass filter over the range of 750 nm to 1100 nm was applied (see section on experiments) and the calculated autocorrelation profiles correspond to that of the filtered SC spectrum. Fast oscillations under the IAC envelope are only observed for very short delays (c.a. $\pm 50$~fs) around the central spike as shown in the figure inset. The rapid decay in the oscillations around the central spike and their absence in the broad pedestal is a consequence of the significant shot-to-shot phase fluctuations of the SC pulses. Notice that the decay of the envelope of the interference pattern from 8 to 2 in the central spike is similar to what is observed for a stationary chaotic light source (see Fig. 1). The coherence time in this case is reduced to $\tau_c = 32$~fs. Note that it is important to compute the IAC profile over an extended delay to observe the full decay of the IAC pedestal from to 2 to 1 since SC spectral components do not overlap in time on average. The AC profile on the other hand shown in (d) is similar to the mean envelope of the IAC. The width of the spike of the AC function is $\approx 150$~fs and it provides an estimation of the average duration of the sub-pulses present underneath the broad average temporal envelope, while the width of the pedestal close to 4 ps corresponds to the average temporal duration of the SC pulses. We also note that the amplitude of the spike in the AC trace has a contrast close to 2:1, similar to what is observed for a chaotic light source.

\section{Experimental Setup}

We next proceeded to the experimental interferometric autocorrelation measurements of SC with characteristics similar to that presented in the numerical simulations section above. The SC is generated by injecting $T_{FWHM} = 210$ fs pulses at 860 nm into the anomalous dispersion regime of a 50~cm long photonic crystal fiber (PCF, Thorlabs NL-2.8-850-02). The  pulses are produced by a mode-locked Ti:Sapphire laser (Spectra-Physics Tsunami) with 80 MHz repetition rate. The zero-dispersion wavelength (ZDW) of the PCF is 850 nm with dispersion slope of 0.48 ps/nm\textsuperscript{2}/km and nonlinear coefficient 47 W\textsuperscript{-1}km\textsuperscript{-1}.

\begin{figure}[htbp]
\centering
\includegraphics[width=\linewidth]{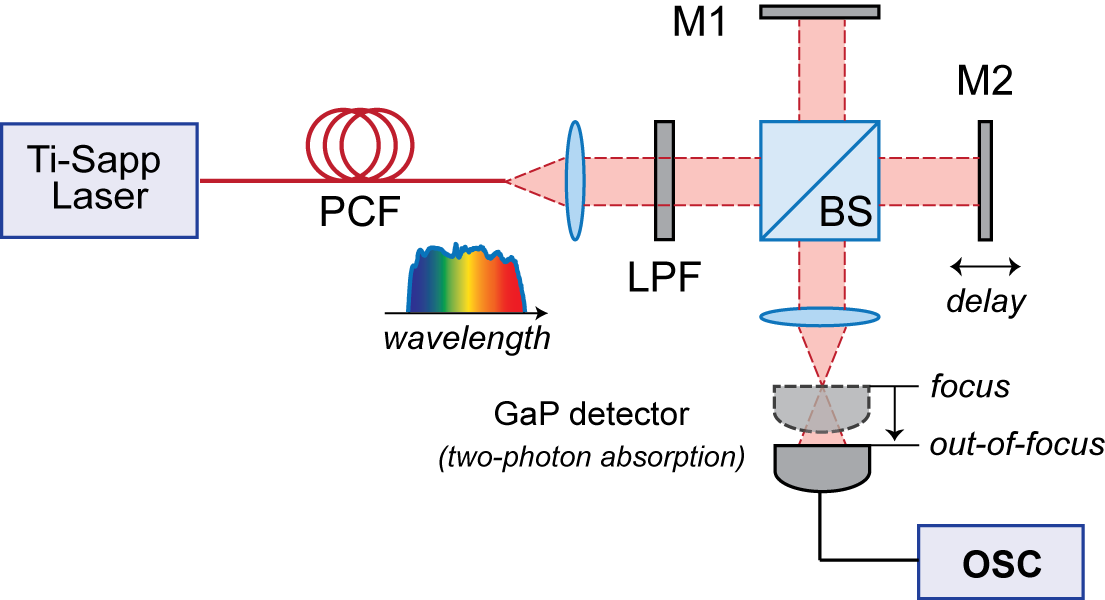}
\caption{Experimental setup. Supercontinuum source generated by injecting $210$~fs pulses from a Ti:Sapphire mode-locked laser through a photonic-crystal fibre (PCF). TPA interferometric autocorrelation measurements are performed in a Michelson interferometer with a fixed (M1) and a moving mirror (M2), beam-splitter (BS), aspheric lens to focus the interfered beams, and a GaP photodetector connected to an oscilloscope (OSC).}
\label{fig:Experimental-setup}
\end{figure}

%% IAC setup
The interferometric autocorrelation setup used here is a Michelson interferometer with a 4~nm precision piezo motor (SmarAct SLC2430) that controls the delay in the scanning arm. Light at the output of the interferometer is focused onto a GaP (transimpedance amplified) photodetector (Thorlabs PDA25K) to perform the two-photon intensity interferometric autocorrelation measurement as shown in Fig. \ref{fig:Experimental-setup}. The photocurrent from the photodetector is measured directly with an oscilloscope as a function of delay in the moving arm. In principle, direct absorption in the GaP photodetector is limited to the 150-550 nm wavelength range, such that two-photon absorption occurs for wavelengths from 550 nm to 1100 nm. However in practice,  impurities in the semiconductor material extends the linear response of the the photodetector and significant direct absorption can still be observed for wavelength above 550 nm. This unwanted linear response of the photodetector is in fact the main challenge in two-photon IAC measurement of broadband SC light as the intensity associated with some of the spectral components of the SC in the normal dispersion regime are relatively weak and the resulting nonlinear absorption may be dominated by direct absorption for those wavelengths. In order to circumvent this problem, we take advantage of the fact that TPA is proportional to the square of the intensity and is thus only observed when the photodetector is placed at the focal plane of the overlapping beams. As the photodetector is displaced from the focal point, nonlinear absorption is significantly reduced up to a point where only linear absorption occurs. Therefore, by subtracting the interferogram recorded when the detector is out of focus at the interferometer output from the interferogram measured when the detector is at the focal plane, one can isolate the true two-photon absorption IAC. Note that when the photodetector is in the out-of-focus position, one needs to ensure that the beam size does not exceed that of the detector such that the full intensity is actually measured.

\section{Results and Discussions}

\begin{figure}[htbp]
\centering
\includegraphics[width=\linewidth]{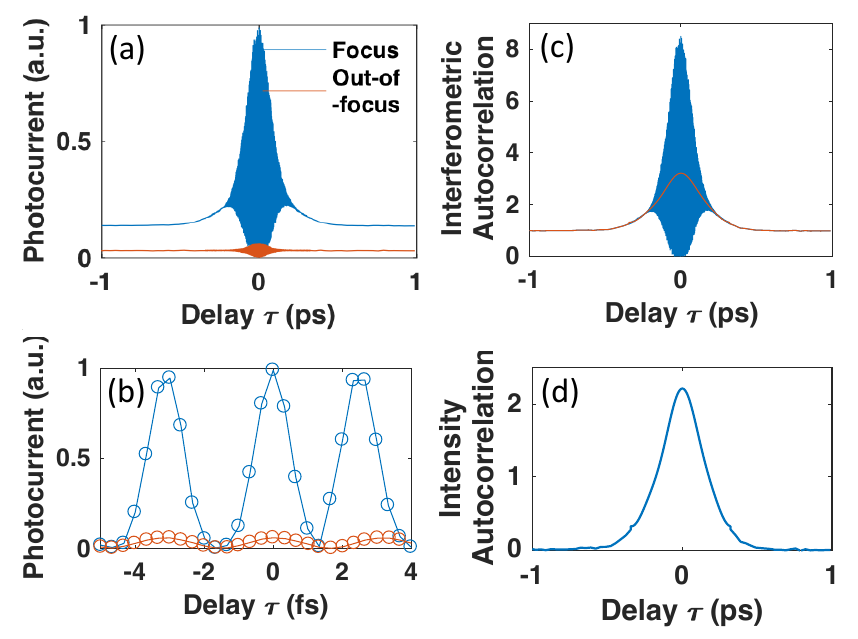}
\caption{Laser characterization from the IAC trace. (a) Measured interferogram at focus plane (blue) and out-of-focus plane (red), with (b) expanded view around zero-delay. (c) The subtracted TPA trace with peak-to-background ratio of 8:1 (blue) and low-pass filtered trace (red). (d) Extracted intensity autocorrelation function.}
\label{fig:Laser-characterization}
\end{figure}

We first tested this two-step approach to measure the interferometric autocorrelation trace of the femtosecond pulses from the Ti:Sapphire laser before injection into the PCF. The measured interferograms recorded in and out of focus for a peak power of 1.7 kW are plotted in Fig. \ref{fig:Laser-characterization}(a) in blue and red, respectively. One can see that there is a high contrast between the two interferograms showing that the nonlinear absorption dominates significantly the linear absorption, where the interferogram measured at the focal plane itself exhibits a peak-to-background ratio close to 8:1, similar to that reported in \cite{Chong2014} and expected for mode-locked pulses (Fig. 1). The expanded view of the interferogram in (b) shows the photocurrent measured around zero delay with each measured delay point indicated by a blue circle. In order to isolate only the TPA contribution, however, the linear contribution must be subtracted following the procedure described above. Figure \ref{fig:Laser-characterization}(c) shows the resulting IAC trace, obtained by normalizing the tail of the subtracted interferogram to 1. The low-pass filtered component of the IAC trace corresponds to the AC function and it is shown in red in the figure. The peak-to-background ratio is close to 3:1 corresponding to a ratio close to 2:0 after background subtraction [see Fig. \ref{fig:Laser-characterization}(d)]. Note that the near-perfect symmetry of the autocorrelation around zero delay is an indication that  the interferometer is properly aligned. The width of the extracted AC profile is $\approx320$ fs corresponding to an effective duration of $\approx210$ fs assuming hyperbolic-secant shape and a deconvolution factor of 0.65 \cite{Sala1980}. The IAC profile on the other hand provides additional information where we can see that interference are not observed across the full delay range for which the interfering pulses overlap, indicating that the pulses are in fact chirped. The chirp is caused by the dispersion in the optical isolator and other optical components of the setup. 

\begin{figure}[htbp]
\centering
\includegraphics[width=\linewidth]{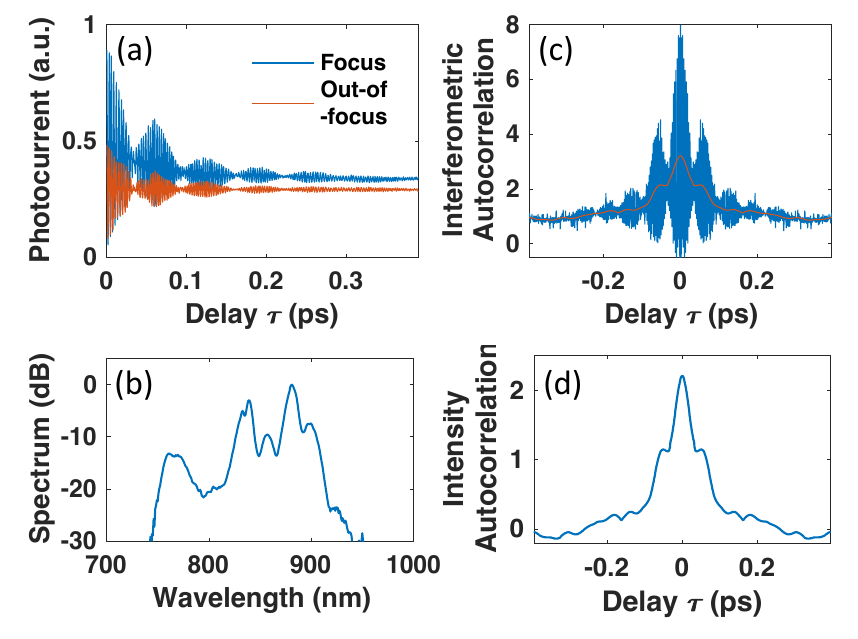}
\caption{Quasi-coherent SC case. (a) Measured interferogram at focus plane (blue) and out-of-focus plane (red). (b) Spectrum measured by an OSA. (c)  IAC trace (blue) and low-pass filtered trace (red). (d) Extracted AC function. Note that the autocorrelation traces for negative delay shown here are obtained by mirroring the measured positive delay traces.}
\label{fig:Coherent-SC}
\end{figure}

We next proceeded to perform TPA IAC measurements of the SC. We first studied the case when the peak power injected into the PCF is relatively low ($P_p \approx 420$ W). At this power level, the spectral broadening is limited from 750 to 950 nm (-30 dB bandwidth) as shown in Fig. \ref{fig:Coherent-SC}(b) with relatively good qualitative agreement with the numerical simulations of Fig. \ref{fig:Sims-Coherent-SC}. We do note a discrepancy on the short wavelength side with a more pronounced dispersive wave component as compared to the numerical simulations, which we attribute to the uncertainty on the dispersion profile provided by the manufacturer and used in the modeling. Figure \ref{fig:Coherent-SC}(a) shows the interferogram recorded in steps of 40 nm step (corresponding to 0.27 fs delay step) by the GaP photodetector placed at the focus (blue) and out-of-focus (red) planes. Note that in order to reduce the overall measurement time and limits potential changes in the coupling conditions, recording was only performed in one direction away from the zero delay. The retrieved IAC profile after subtracting the linear absorption contribution is shown in (c) and we remark very good agreement with that of the simulations shown in Fig. 2(c). Specifically, the measured peak-to-background ratio is close to 8:1 and we observe multiple side-lobes with fast oscillations visible across the full duration of the SC pulses, indicative of multiple phase-stable sub-structures under the temporal envelope. The corresponding coherence time is $300$~fs, also in agreement with the simulations. The extracted intensity autocorrelation trace extracted by low-pass filtering is plotted in (d) and it has a temporal width of the same order as that in the simulations with the expected peak-to-background ratio of 2:0. 

\begin{figure}[htbp]
\centering
\includegraphics[width=\linewidth]{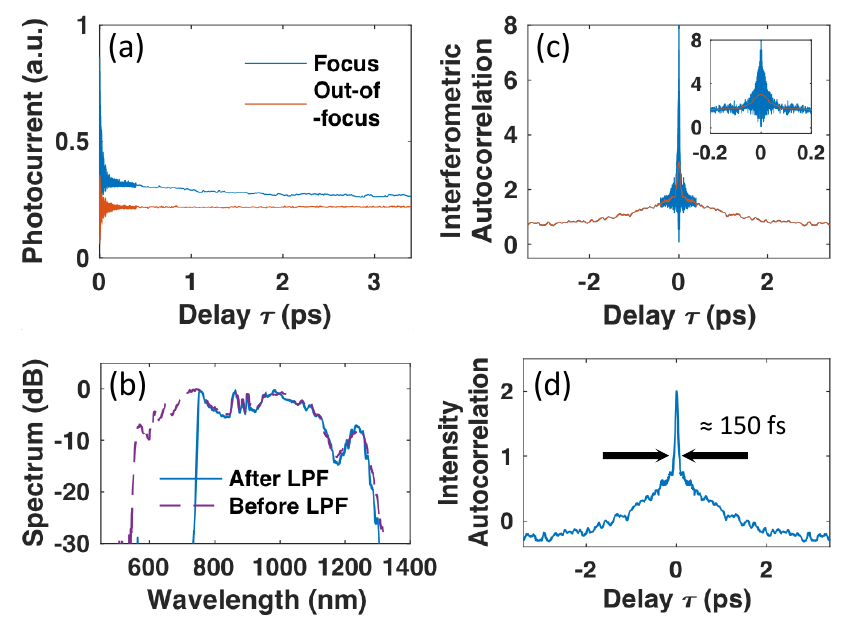}
\caption{Quasi-incoherent SC case. (a) Measured interferogram at focus plane (blue) and out-of-focus plane (red). (b) Spectrum measured by an OSA. (c)  IAC trace (blue) and low-pass filtered trace (red). (d) Extracted AC function. Note that the autocorrelation traces for negative delay shown here are obtained by mirroring the measured positive delay traces.}
\label{fig:Incoherent-SC}
\end{figure}

We subsequently increased the peak power to $P_p = 18~kW$ leading to an increase in the SC bandwidth with the average spectrum plotted in Fig. \ref{fig:Incoherent-SC}(b) now spanning from c.a. 550 nm to 1300 nm (-30 dB bandwidth) in relative good agreement with the numerical simulations shown in Fig. \ref{fig:Sims-Incoherent-SC}(b). The discrepancy is again probably caused by the uncertainty on the dispersion profile. In order to avoid significant linear absorption that still occurs in the GaP detector for wavelengths up to 700 nm and even beyond, we use a long-pass filter with cut-on wavelength at around 750~nm. Note also that spectral components above 1100 nm do not contribute to the two-photon absorption signal and are thus filtered out by the detector. The measured interferograms in and out-of focus are plotted in Fig. \ref{fig:Incoherent-SC}(a) while (c) and (d) show the temporal IAC profile and extracted AC, respectively, measured for a total delay of 3.9 ps. Note that in order to minimize possible variations in the injected peak power caused by slow thermal effects at large values during the acquisition of the IAC, the measurement time was reduced by using different precision in the scanning mirror step for small and large delays from the zero-delay position. Specifically, for delays below 400 fs, corresponding to the span of the central spike a step of 30 nm (corresponding to a 0.2 fs resolution) was used, while a larger step of $1.5\thinspace\mu$m (corresponding to a 10 fs resolution) was used for longer delays corresponding to the broad pedestal part of the profile. Fast oscillation are only observed around the central spike with a peak-to-background ratio close to 8:2, confirming the chaotic light nature of the SC generation process. The coherence time estimated from the IAC profile is 75 fs, about twice that obtained from the numerical simulations and consistent with the fact that the simulated spectra have an increased bandwidth. The width of the spike of the AC function here (giving the estimation of the average duration of sub-pulses underneath the broader SC temporal envelope) is 120 fs, while the width of the pedestal gives the average total span of the SC pulses, close to 4 ps. These values differs slightly from that obatined in the numerical simulations pointing again towards a slightly different fiber dispersion profile. We also note that the amplitude of the spike in the AC trace has the expected contrast close to 2:1 typical of a chaotic light source.

\section{Conclusions}

We have shown that low-cost and simple IAC measurement of broadband supercontinuum can be achieved using a semiconductor GaP photodetector based on two-photon absorption. We have demonstrated that the linear absorption due to the impurities of the semiconductor material of the photodetector can be eliminated using in and out of focus measurements, allowing direct access to the sheer IAC profile. The interference pattern for a coherent SC extends over the entire IAC trace, while for a quasi-incoherent SC it only extends over a few tens of femtosecond temporal delay corresponding to only a fraction of the SC average duration. The delay over which the interference pattern is observed provides an estimation of the coherence time while the envelope also provides information about the presence of sub-pulses. The extracted AC trace on the other hand has the characteristics of mode-locked pulses in the case of a low power SC while it has the characteristics of a chaotic stationary light source around the zero-delay for increased power. We anticipate that the simplicity of the technique may be very useful in studying the coherence properties of light sources.

\section{Funding Information}

We acknowledge the support from the Academy of Finland (grant 298463)

\bibliographystyle{abbrv}
\bibliography{Mendeley_CoherenceSC}
\end{document}